\def\slash#1{\setbox0=\hbox{$#1$}#1\hskip-\wd0\dimen0=5pt\advance
\dimen0 by-\ht0\advance\dimen0 by\dp0\lower0.5\dimen0\hbox
to\wd0{\hss\sl/\/\hss}}

\documentstyle[aps,epsfig,prl]{revtex}

\renewcommand{\prl}[3]{Phys.\ Rev.\ Lett.\     {\bf  #1},  #2    (#3)}

\begin{document}
\preprint{BARI-TH/98-317}
\title{Multiscale Analysis of Blood Pressure Signals}
\author{A. Marrone, A.D. Polosa, and G. Scioscia }
\address{Dipartimento di Fisica Universit\`a di Bari
 and Sezione INFN di Bari, \\
 Via Amendola 173, I-70126 Bari, Italy}
\author{S. Stramaglia}
\address{Istituto Elaborazione Segnali ed Immagini,\\
Consiglio Nazionale delle Ricerche,\\
Via Amendola 166/5, I-70126 Bari, Italy}
\author{A. Zenzola}
\address{Dipartimento di Scienze Neurologiche e 
	Psichiatriche Universit\`a di Bari, \\ 
	Piazza Giulio Cesare 11, I-70100 Bari, Italy}

\date{October 1998}
\maketitle
\begin{abstract}
We describe the multiresolution wavelet analysis of blood pressure
waves in vasovagal syncope affected patients compared
with healthy people one, using Haar and Gaussian bases. 
A comparison between scale-dependent and scale-independent measures
discriminating the two classes of subjects is made. 
What emerges is a sort of equivalence between these 
two methodological approaches, that is both methods reach the
same statistical significance of separation between the two classes.
\end{abstract}
\vspace*{1.4truecm} 
In recent years biological time series have been considered in the more
general framework of {\it fractal functions}. Accordingly, the analysis
tools commonly used for fractal functions, have been applied to study
physiological time series, see e.g.~\cite{nature}.
A major approach to such problems is based on
the {\it wavelet} transform, a technique which has proved to be well
suited for characterizing the scaling properties of fractal objects even
in presence of low-frequency trends~\cite{me}.

In particular, the regulation of the cardiac rhythm, has been recently
investigated
in two very interesting papers aiming at providing
means of diagnosis of heart disease. The interbeat
interval records for healthy and sick subjects have been studied 
by wavelet analysis, which can appropriately treat the non-stationarity
of these signals. In Ref.~\cite{Th98} a scale dependent measure, the root-mean
square   of
the wavelet coefficients $\sigma_w (s)$ 
at a particular scale $s$, has shown to be able to sharply
discriminate between healthy and sick subjects. In Ref.~\cite{hestanley} it was
observed that scale-dependent measures may reflect characteristics
specific to the subject or to the choice of the wavelet basis; a
scale-independent measure extracting the exponents characterizing
the scaling of the partition function of wavelet coefficients
was then proposed,  and its
performance in detecting heart disease was excellent.
The scaling exponents were already studied in Ref.~\cite{Th98} for the second
wavelet moments while in Ref.~\cite{hestanley} they  are calculated for
arbitrary moments.
It seems likely to us that the approaches based on 
scale-dependent measures and that based on
scale-independent ones should be considered as qualitatively {\it equivalent}.
Indeed some pathological conditions may alter the cardiac dynamics at a
specific scale or range of scales, while the scaling behaviour of the
dynamics of cardiac rhythm regulation should be universal for subjects
belonging to the same class. An important problem is how the choice of
the wavelet basis influences the results of the analysis. Moreover it
is interesting to check whether the same kind of analysis can be used to
study {\it other} physiological time series and pathologies. 

In this work we study 
the temporal series of the systolic blood pressure
waves maxima in nine healthy subjects and ten 
subjects showing a pathology known 
as vasovagal syncope.
We perform a wavelet analysis of these time series and consider both
the scale-dependent and the scale-independent measures above described.
To our knowledge this is the first time wavelets are used to study
blood pressure waves signals. We performed the analysis using two
different wavelet bases, namely the Haar basis and the third derivative
of the Gaussian one (TDG). The main difference between these two bases is that
the former is able to remove only zero order trends, while the latter
is insensible to higher polynomial trends.
We find that the Haar basis is, with respect to the data set at hand,
well suited for scale-dependent measures i.e. measures of $\sigma_w(s)$. 
Indeed using these wavelets we
find an evident separation among healthy and sick subjects at a
particular scale $s=32$: this separation is missing when the TDG
is used. On the other hand, using the TDG
leads to a separation with respect to scaling exponents 
measures which is quite less significant when the Haar basis is used.
Interestingly we found that the
statistical confidence of separation in the scale-dependent parameter 
(obtained using Haar wavelets) is very close to that obtained by the
scale-independent parameter i.e. the scaling exponent (using the TDG). 
Since both methods have 
reached the  same degree of  separation
between the two classes, it remains to be understood whether 
this coincides with the intrinsic
degree of separability of the data set here considered.

Vasovagal syncope is a sudden, rapid and reversing 
loss of consciousness, due to a reduction of cerebral blood 
flow~\cite{Ka95} attributable  to a dysfunction
of the cardiovascular control,
induced by that part of the   
Autonomic Nervous System (ANS) that regulates 
the arterial pressure~\cite{Ka95,Wo93}.
In normal conditions the arterial pressure is maintained 
at a constant level by the existence of a negative feed-back 
mechanism localised in some nervous centres of the brainstem. 
As a  consequence of a blood pressure variation, 
the ANS is able to restore the haemodynamic situation acting
on  heart and vases, by means of two efferent pathways,
the vasovagal and sympathetic one, the former acting in the sense 
of a reduction of the arterial pressure, the latter in the opposite
sense~\cite{Gr66}.
Vasovagal syncope consists of an abrupt fall of blood pressure  
corresponding to an acute haemodynamic
reaction produced by a sudden change in the activity of the ANS 
(an excessive enhancement of vasovagal outflow or a sudden decrease 
of sympathetic activity)~\cite{Ka95}.

Vasovagal syncope is a quite common clinical problem and 
in the~$50\%$ of patients it is non diagnosed, being 
labelled as syncope of unknown origin, i.e. not necessarily 
connected to a dysfunction of the ANS action.\cite{Wo93,K95b,Ru95}. 
Anyway, a rough diagnosis of vasovagal syncope 
is practicable~\cite{K95b,Ka94}  
with the help 
of the head-up tilt test (HUT)~\cite{Ke86}.
During this test the patient, positioned on a self-moving table, 
after an initial rest period in horizontal  position, 
is suddenly brought in vertical  position. 
In such a way the ANS registers a sudden stimulus of reduction 
of arterial pressure due to the shift of blood volume 
to inferior limbs. A badly regulated response to this stimulus 
can induce syncope behaviour.

According to some authors, the positiveness of HUT means 
an individual predisposition toward vasovagal syncope \cite{Gr96}.
This statement does not find a general agreement because of 
the low reproducibility of the test~\cite{Ru96} in the same patient
and the extreme variability of the sensitivity in most 
of the clinical studies~\cite{Ka94}.
For this reason a long and careful clinical observation period is
needed to establish with a certain reliability whether 
the patient is affected
by this syndrome.
What we want to stress here is that, from a clinical standpoint, there is 
{\it not} a neat way of discriminating between healthy and 
syncope-affected subjects, 
while, in the case of heart disease, studied in~\cite{Th98,hestanley},
there is always a very clear clinical picture.
For this reason in last years a large piece of work has been devoted to 
the investigation of signal patterns that could characterise
syncope-affected patients. This has been done especially
by means of Fourier analyses of  arterial pressure and 
heart rate which have not shown to be successful for this purpose~\cite{Wa61}.
 
The temporal behaviour of blood pressure is the most 
clinically relevant aspect to study vasovagal syncope since it is the
result of the combined activity of ANS on heart and vases.
Therefore we extract blood
pressure wave maxima from a recording period  twenty minutes long (which
is the better we can do for technical reasons). 
During this time the following biological signals of the subject
are recorded: E.C.G. (lead  D-II), E.E.G., the thoracic breath,
the arterial blood pressure (by means of a system {\em finapres~Ohmeda~2300}
Eglewood co. USA, measuring from the second finger of the left hand).

We denote $\{P_i\}$ the time series of systolic 
pressure maxima. The coefficients
of the discrete wavelet transform at scale $s$ are given by:
\begin{equation}
W_s(n)=s^{-1}\sum_{i=1}^{M} P_i \psi((i-n)/s),
\end{equation}
where $\psi$ is the generating wavelet, $M$ is the number of points in
the time series (we have $M=2^{10}$), 
$n$ is the point for which the coefficient is
calculated. The scale-dependent measure proposed in~\cite{Th98} corresponds to
evaluate
the root-mean square of wavelet coefficients at fixed scales.
The scale-independent measure deals with the sums of the moments of the
wavelets coefficients 
\begin{equation}
Z_q(s)=\sum_{n} |W_s(n)|^q~~,
\end{equation}
where the sum is only over the maxima of $|W_s|$. One can show that
$Z_q$ scales as:
\begin{equation}
Z_q(s)\sim s^{\tau(q)}.
\end{equation}

The exponents $\tau(q)$, especially for $q=2$ and $q=5$, were found to
provide a robust degree of separation in the case of heart disease
diagnosis~\cite{hestanley}.

Firstly we discuss the results we obtained on the data set here
considered by the evaluation of the r.m.s. of wavelet coefficients.
In Fig.~1(a) the r.m.s. of the Haar wavelets coefficients are plotted
versus the scale, for both sick and healthy subjects, while in Fig.~1(b)
the same quantities are plotted in the case of the TDG
 basis. One can see that in the Haar case an evident
separation between healthy and sick subjects holds at the scale $s=32$:
healthy subjects have greater fluctuations in the wavelet coefficients.
We performed the WMW (Wilcoxon-Mann-Whitney) test to check the hypothesis
that the two kinds of samples, positive and control subjects, have been
drawn from the same continuous distribution function. The WMW test gives a
$3.5\times 10^{-3}$ probability to the above cited hypothesis, i.e. the
statistical hypothesis is rejectable at the level of significance of
$1\%$. On the other hand, using the TDG as
the wavelet basis does not lead to separation at any scale.
Therefore a scale-dependent measure can   highlight an   
evident separation at a particular scale but the result depends on the
wavelet basis one uses.

Let us now turn to consider scaling exponents measures. We have
calculated the partition functions  $Z_q(s)$ using both Haar wavelets
and the TDG basis. A measure of the
exponents $\tau (q)$ can then be obtained through log-log plots of $Z_q$
versus $s$. 
In the case of the TDG,
the log-log plots of $Z_q$
versus $s$ showed a neat scaling 
behaviour: in Fig.~2 the $q=1$ case, the most significant 
with our data, is shown. 
In the case of Haar wavelets, the log-log plots show some
curvature (see Fig.~2), but calculating linear correlation coefficients
we discover that it still makes sense to evaluate $\tau(q=1)$ exponents.
For the moment let's refer to the TDG case.
We found that the exponent $\tau (q=1)$ acts as a
discriminating parameter between healthy and sick subjects, while 
exponents for the other values of $q$ did not succeed in obtaining
equally convincing results. Healthy subjects  have lower $\tau (q=1)$ 
values than syncope affected ones (see. Fig. 3).
By WMW test, the probability that the values
of the exponents found for the two classes of subjects, healthy and
sick, were sampled from the same continuous distribution was estimated
$4.5\times 10^{-3}$, a level of significance very close to the one found
in the case of the scale-dependent measure.
On the other hand, considering the $\tau(q=1)$ as computed in the Haar basis,
we find that the latter probability value grows of about one order of
magnitude reaching a value of $2.1 \times 10^{-2}$.

In Fig.~3 we have shown the points corresponding to the $19$ subjects
under consideration in the $\sigma_w -\tau$ plane, where the coordinates
correspond to the measured quantities $\sigma_w (32)$ (by Haar wavelets) and
$\tau(q=1)$ (by the TDG basis). It is
evident   
that the two measures separate, {\it at the same degree}, the two classes.

We observe that it is reasonable that  the
Gaussian basis is more effective in detecting the scaling behaviour
of these time series with respect to the Haar basis.
On the other hand
the same degree of separation is obtained by
Haar wavelets at a given scale, while the TDG
seems insensible to the single scale features. It follows that
these two kinds of measures are going in the same direction
rather  than excluding each other.
A very careful analysis in Ref.~\cite{IITh98} shows that in the case 
of diverse
heart pathologies, scale-dependent measures, namely measures of 
$\sigma_w (s)$ at a particular scale $s$, outperform measures of scaling
exponents. Due to the size of our data set, we may encounter problems in
reproducing the kind of analysis performed in Ref.~\cite{IITh98}, but we
look forward to investigate this aspect. In Ref~\cite{IITh98} is also stressed
that baroflex modulations of sympathetic and parasympathetic tone lie
in a frequency range which corresponds to the scale $s=32$ which is, 
also for us, the best discriminating between controls and positives. 

We are, at the moment, not able to provide the
physiological explanation of the phenomena here described. However these
results might be useful to get a better understanding of
the very complicated vasovagal syncope pathology.

Some conclusions are in order. We analysed by wavelets blood pressure
signals from healthy subjects and subjects positive to vasovagal syncope
pathology. We evaluated two quantities, one depending on a fixed scale
and a scaling exponent, which have been recently proposed as
diagnosis tools for heart disease. We have shown that both the measures
succeed in separating the two classes within the same degree of
significance. 
We are working to have
longer records and an enlarged number of positives and controls
so as to refine our analysis. At the moment we are 
aware to be far from being able to propose an alternative 
diagnostic tool. This would be very useful in consideration of the 
particular difficulty that the clinical diagnosis of 
vasovagal syncope still presents.

\noindent
The authors are grateful to Doc. M.~Osei~Bonsu for giving us 
the possibility to access at the not elaborated 
blood pressure data. A.D.P. Thanks the group-IV of INFN and in particular 
Prof. R.~Gatto for supporting his permanence at CERN-Geneva.
The authors also thank an anonymous referee whose suggestions improved
the presentation of this work.

\newpage


\vspace*{4truecm}
\begin{figure}
\centerline{\epsffile{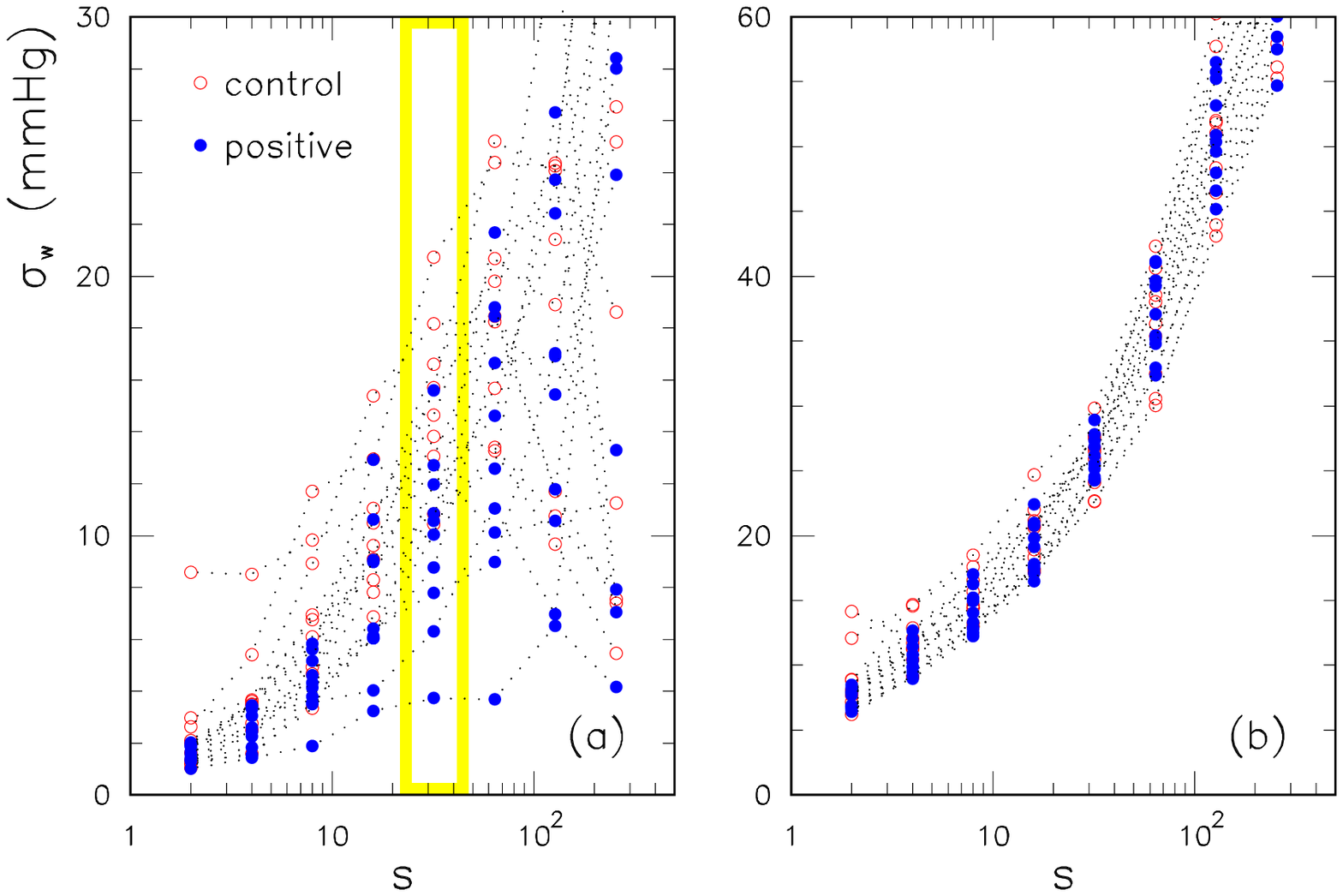}}
\noindent
{\bf Fig. 1} - {Standard deviations of the wavelet coefficients 
                of the systolic  pressure in syncope-affected 
		patients (positives) and healthy people (controls)
                drawn both in Haar basis (a) and Gaussian basis (b).
		Note the  evident separation among positives
		and controls at $s=32$ in (a). This separation is
                completely lost in (b). We believe that the discrimination
		pattern is not as sharply evident as in Ref.~[3] due to
		the restrict temporal extension of our data set.
		We use different $\sigma_w$ ranges in (a) and (b)
		just in order to have
		a best visual impact of figures.}
\end{figure}

\newpage 

\vspace*{4truecm}

\begin{figure}
\centerline{\epsffile{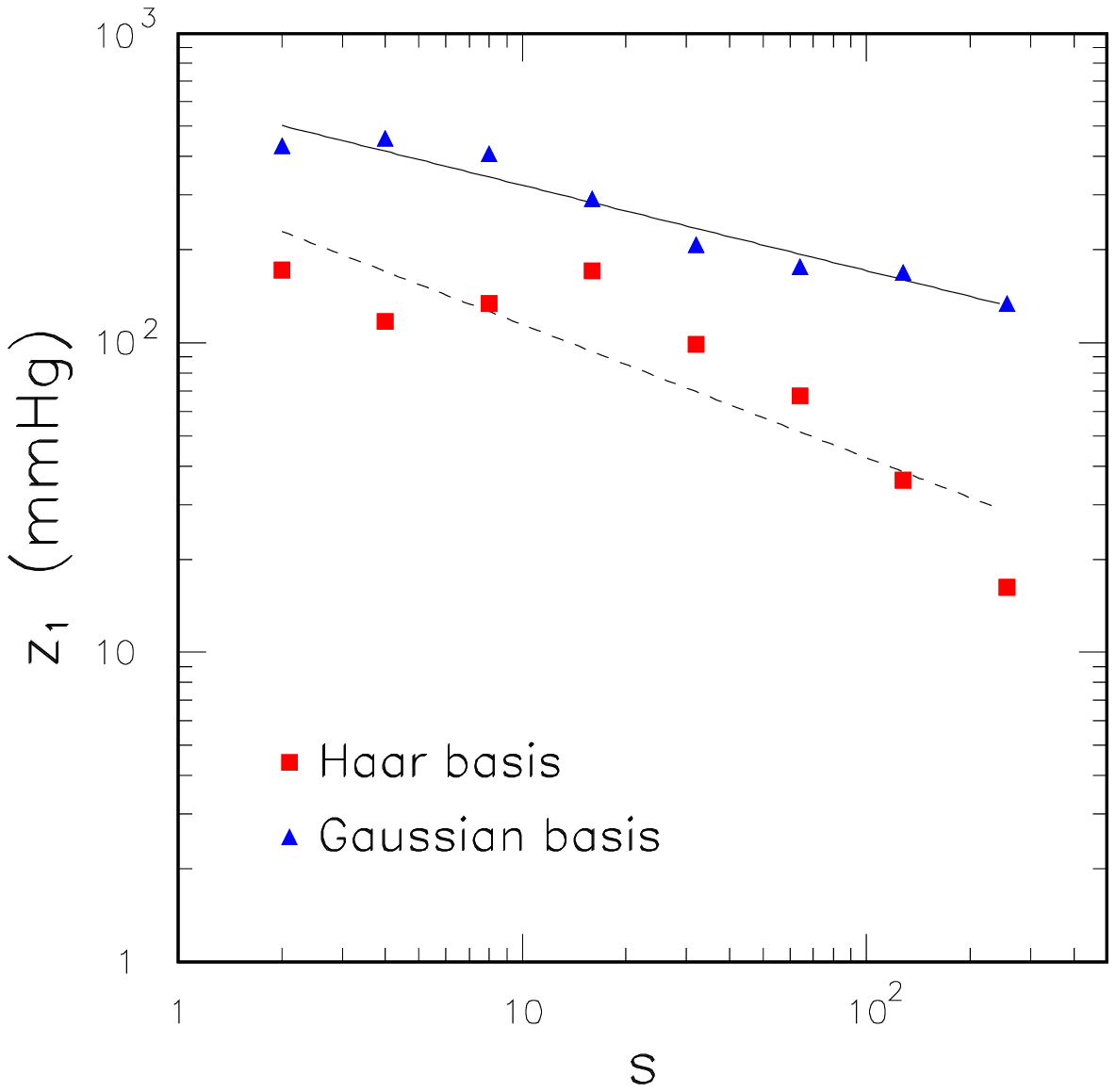}}
\noindent
{\bf Fig. 2} - {Log-log plots of $Z_1(s)$~vs.~$s$ drawn in Haar basis  
		 and in Gaussian basis for a subject.
                 Analogous plots have been obtained for the
		 other subjects we have examined.   
		 The  scaling behaviour
		 is evident only in Gaussian basis, while the points
		 from the Haar basis are not as well linearly fitted.}
\end{figure}

\newpage 

\vspace*{4truecm}

\begin{figure}
\centerline{\epsffile{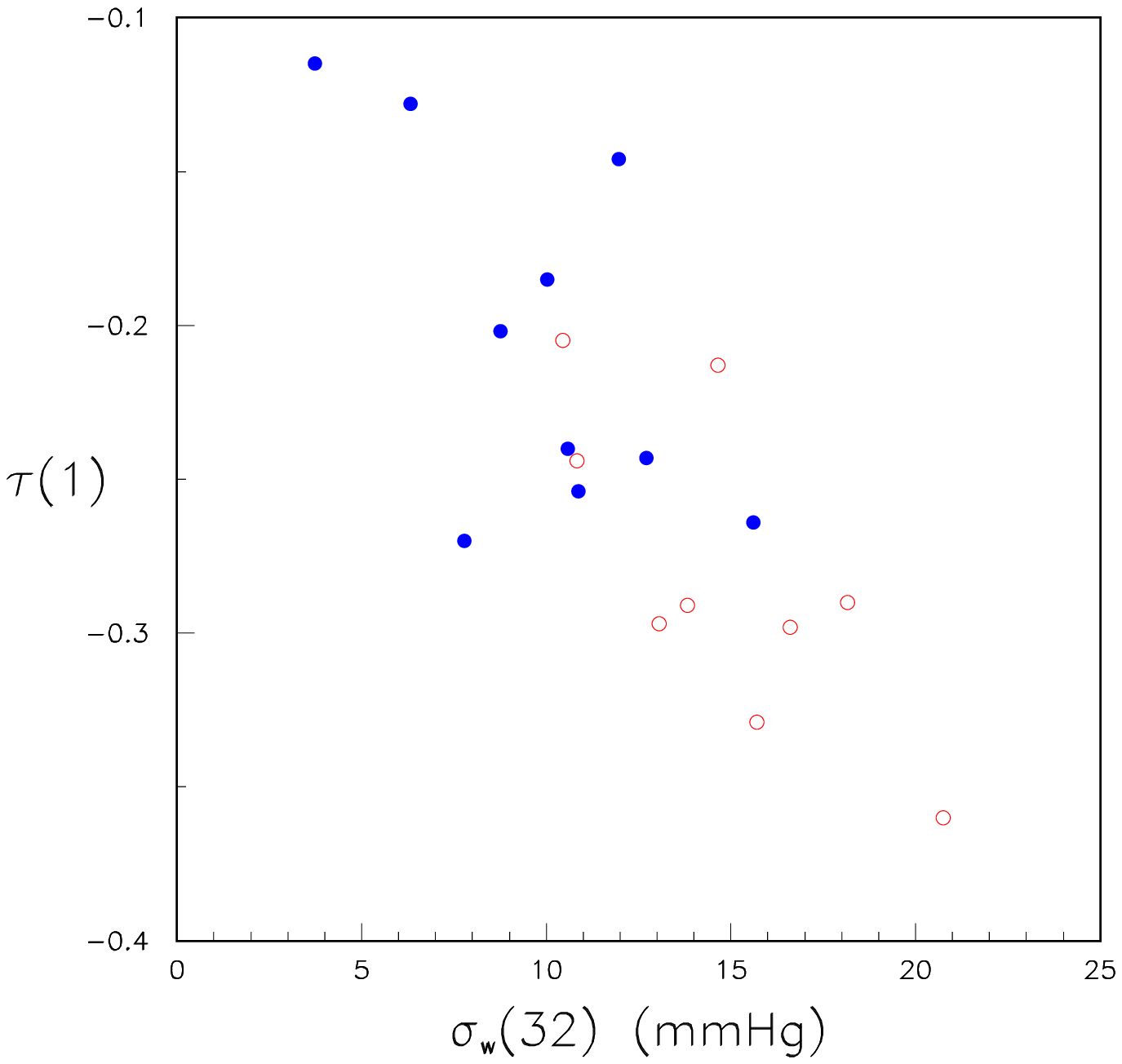}}
\noindent
{\bf Fig. 3} - {$\sigma_w(32)$-$\tau(1)$ plot
                 i.e. the Haar wavelet coefficient fluctuation
		 at the scale $s=32$ vs. the scaling exponent of
		 Eq. (3) in Gaussian basis. $\bullet$ is referred to
		 syncope-affected patients, $\circ$ is referred to healthy
		 subjects. Projecting the points laying in the $\sigma-\tau$
		 plane on $\sigma$ and $\tau$ axes we obtain two
		 separation patterns between positives and controls which
		 are quite similar from a statistical point of view (see
		 the WMW analysis in the text).}
\end{figure}

\end{document}